%% file: main.tex
\documentclass[conference]{IEEEtran}
\IEEEoverridecommandlockouts
\usepackage{cite}
\usepackage{amsmath,amssymb,amsfonts}
\usepackage{algorithmic}
\usepackage{graphicx}
\usepackage{textcomp}
\usepackage{xcolor}
\usepackage{threeparttable}
\usepackage{multirow}
\usepackage{colortbl}
\usepackage{color}
\usepackage{tabularray}
\usepackage{fancyhdr}

\def\BibTeX{{\rm B\kern-.05em{\sc i\kern-.025em b}\kern-.08em
    T\kern-.1667em\lower.7ex\hbox{E}\kern-.125emX}}

\definecolor{Orange}{rgb}{1.0, 0.5, 0.0}
\definecolor{Purple}{rgb}{0.5, 0.0, 0.5}
\definecolor{MagicMint}{rgb}{0.71,0.88,0.804}
\definecolor{Festival}{rgb}{1.0,0.898,0.6}
\definecolor{BadColor}{rgb}{1.0,0.294,0.294}

\newcommand{\blue}[1]{{\color{black}#1}}
\newcommand{\orange}[1]{{\color{black}#1}}
\newcommand{\purple}[1]{{\color{black}#1}}

\begin{document}


\title{Fault Tolerant Design of IGZO-based Binary\\Search ADCs}


\fancypagestyle{firstpage}{%
    \fancyhf{}  
    \renewcommand{\headrulewidth}{0pt}  
    \renewcommand{\footrulewidth}{0pt}
    \fancyhead[C]{Accepted for publication at the 27th International Symposium on Quality Electronic Design (ISQED'26), April 8-10, 2026.}
}

\newif\ifanonym
\anonymfalse 

\ifanonym
\author{}
\else
\author{\IEEEauthorblockN{Paula Carolina Lozano Duarte}
\IEEEauthorblockA{
Karlsruhe Institute of Technology\\
Karlsruhe, Germany\\
paula.duarte@kit.edu}
\and
\IEEEauthorblockN{Sule Ozev}
\IEEEauthorblockA{Arizona State University\\
Tempe, USA\\
Sule.Ozev@asu.edu}
\and
\IEEEauthorblockN{Mehdi Tahoori}
\IEEEauthorblockA{
Karlsruhe Institute of Technology\\
Karlsruhe, Germany\\
mehdi.tahoori@kit.edu}
}
\fi

\maketitle
\thispagestyle{firstpage}
\input{section/0_Abstract}
\input{section/1_Introduction}
\input{section/2_Background}
\input{section/3_Methodology}
\input{section/4_Results}
\input{section/5_Conclusion}

{\small
\section*{Acknowledgment}
This work has been supported by the European Research Council (ERC) (Grant No. 101052764) and the KIT International Excellence Fellowship.
}

\bibliographystyle{IEEEtran}
\bibliography{IEEEabrv,references}

\end{document}

%% file: section/0_Abstract.tex
\begin{abstract}
Thin-film technologies such as Indium Gallium Zinc Oxide (IGZO) enable Flexible Electronics (FE) for emerging applications in wearable sensing, personal health monitoring, and large-area systems. 
Analog-to-digital converters (ADCs) serve as critical sensor interfaces in these systems. Yet, their vulnerability to manufacturing defects remains poorly understood despite \orange{unipolar technologies'} inherently high defect densities and process variations compared to mature CMOS technologies.
\purple{We present a hierarchical fault injection framework to characterize defect sensitivity in Binary Search ADCs implemented in \orange{n-type only} technologies.} 
Our methodology combines transistor-level defect characterization with system-level fault propagation analysis, enabling efficient exploration of both single and multiple fault scenarios across the conversion hierarchy.
\purple{The framework identifies critical fault-sensitive circuit components and enables selective redundancy strategies targeting only the most sensitive components.}
\purple{The resulting defect-tolerant designs} improve fault coverage from 60\% to 92\% under single-fault injections and from 34\% to 77.6\% under multi-fault injection, while incurring only 4.2\% area overhead and 6\% power increase.
\orange{While validated on IGZO-TFTs, the methodology applies to all emerging unipolar technologies.}
\end{abstract}

\begin{IEEEkeywords}
Flexible Electronics, IGZO-TFT, Fault Injection, Binary Search ADC, Defect-Tolerant Design, Hierarchical Simulation.
\end{IEEEkeywords}

%% file: section/1_Introduction.tex
\section{Introduction}\label{sec:intro}

Flexible electronics (FE) have emerged as a transformative platform for conformable, lightweight, and cost-effective systems where rigid silicon solutions are impractical~\cite{Zhu:IGZO2021, circuits_A-IGZO_Mallory}. Indium Gallium Zinc Oxide (IGZO) thin-film transistors (TFTs) dominate as the active devices in FE due to their low-temperature process compatibility, uniformity across large substrates, and adequate performance for low-frequency applications~\cite{Pan:IGZOTFT2024}.

Analog and mixed-signal circuits are critical building blocks in FE, bridging sensing domains with digital processing~\cite{afentaki:date26:mixed_signal}. Analog-to-digital converters (ADCs) in particular determine signal fidelity in wearable health monitors, environmental sensors, and human-machine interfaces~\cite{Gao:FlexibleWearableSensing, afentaki2025islped, shatta2025ICCAD}. 
\orange{However, emerging thin-film technologies based on unipolar (n-type only) transistors—including IGZO, Indium Tin Oxide (ITO), Gallium Nitride (GaN), and negative capacitance FETs (NCFETs)—face reliability challenges uncommon in mature CMOS: high defect densities, leakage-dominated power due to poor subthreshold behavior, lack of complementary devices, and significant process-induced variations~\cite{azam:analogdefect, BinuSurvey}.}

Although extensive fault modeling exists for CMOS analog circuits~\cite{XiaChebyshev, SubrahmaniyanAdaptive, LiuHilbert, FarooqNLARX}, the susceptibility of analog building blocks in FE remains largely uncharacterized. 
The material characteristics and fabrication process of IGZO-TFTs necessitate fault modeling approaches that capture technology-specific failure mechanisms and their propagation across circuit hierarchies.

This work addresses the gap between conventional fault simulation methodologies and the specific requirements of FE. 
Our objective is to integrate a comprehensive fault sensitivity analysis into the FE design flow, allowing fault resilience to be treated as a first-class design metric alongside traditional performance, power, and area constraints. 

We focus on Binary Search ADC architecture, which offers significant advantages for resource-constrained FE applications. Unlike Flash converters requiring large parallel comparator arrays, Binary Search ADCs perform sequential voltage range subdivision, dramatically reducing component count, silicon area, and static power consumption~\cite{Lozano:aspdac25:BinCoDesign}. This efficiency makes them particularly attractive for battery-operated wearable devices operating at low to moderate sampling rates (sub-100 Hz) typical of biomedical signal acquisition. 
However, the \purple{cascaded} decision-making structure of the Binary Search conversion introduces unique fault propagation characteristics. Unlike parallel topologies, where comparator failures affect isolated output codes, defects in \purple{cascaded} architectures can \purple{propagate} through multiple conversion stages, potentially corrupting entire conversion sequences. 
This hierarchical fault propagation mechanism demands careful analysis to understand how localized defects manifest at the system level.

Our methodology extends hierarchical fault simulation frameworks~\cite{aksoy2021hierarchical} to accommodate FE characteristics, evaluating both isolated and concurrent faults, reflecting the clustered nature of manufacturing defects~\cite{modala2024increasing, aksoy2021hierarchical}. 
\purple{We demonstrate the approach on a 3-bit IGZO Binary Search ADC—a resolution representative of resource-constrained FE applications~\cite{Papadopoulos:C-2CSAR, Lozano:aspdac25:BinCoDesign}—via systematic transistor-level defect injection to identify vulnerability patterns.}
\purple{Based on this fault sensitivity analysis, we develop fault-tolerant ADC designs employing topology modifications and selective redundancy targeted at the most critical components.}
\orange{The hierarchical nature of the methodology enables straightforward extension to higher-resolution ADCs: for an N-bit Binary Search converter, the fault space scales linearly with N (requiring N comparison stages), and the selective redundancy strategy remains applicable as critical vulnerability points concentrate in early conversion stages regardless of resolution~\cite{Lozano:aspdac25:BinCoDesign}.}

Experimental results show substantial improvements in robustness: single-fault coverage increases from 60\% to 92\%, and multi-fault coverage rises from 34\% to 77.6\%, with only 4.2\% area and 6\% power overhead. These results confirm that targeted architectural robustness can significantly enhance reliability under the constraints of FE.

\textbf{Main contributions of this work are as follows:}
\begin{itemize}
\item A hierarchical fault simulation framework supporting multi-fault scenarios in IGZO analog circuits, capturing technology-specific failure modes and fault propagation.
\item Systematic fault sensitivity analysis of Binary Search ADCs, identifying critical vulnerability points and quantifying their impact on conversion accuracy.
\item \purple{Fault-tolerant} ADC designs optimized for IGZO implementation, validated across single- and multi-fault scenarios with hardware cost assessment.
\end{itemize}

%% file: section/2_Background.tex
\section{Background}\label{sec:background}

\subsection{Flexible Electronics and IGZO Technology}\label{sec:FE}

Flexible electronics (FE) encompasses devices built on deformable substrates that maintain operational integrity while being bent, stretched, or twisted. This mechanical adaptability facilitates deployment on non-planar surfaces, including biological tissues, textiles, and soft-robotic components, positioning FE as a particularly suitable technology for wearable health monitoring systems, conformable sensing devices, and interactive human-machine interfaces~\cite{Gao:FlexibleWearableSensing, afentaki2025islped}.
In contrast to conventional rigid silicon-based technologies, FE devices must operate in environments demanding mechanical compliance, cost-effectiveness, and resilience to physical stress. To achieve these characteristics, FE systems are typically manufactured on substrates such as polyimide, plastic polymers, or thin metallic foils, necessitating modifications to standard semiconductor fabrication processes.

Indium-gallium-zinc oxide (IGZO) constitutes a key enabling material in FE, serving as a wide-bandgap semiconductor in thin-film transistor (TFT) technologies. 
IGZO-TFTs exhibit relatively high electron mobility, good uniformity, and compatibility with low-temperature deposition processes, making them well-suited for large-area, low-power applications~\cite{Zhu:IGZO2021, Pan:IGZOTFT2024}.

The absence of complementary (p-type) devices represents a technological limitation that frequently results in increased static power consumption and necessitates alternative design strategies, such as dynamic logic or pseudo-CMOS configurations, for implementing logic and analog circuits.
Despite these constraints, IGZO-based FE circuits have demonstrated the feasibility of integrating essential analog and mixed-signal building blocks. 
Prior research has reported implementations of operational amplifiers~\cite{Zysset:opamptft, Meng:opampigzo}, signal conditioning stages~\cite{ Garripoli:analogfrontendigzo, lozano:AKAN}, and various analog-to-digital converter (ADC) architectures~\cite{Papadopoulos:C-2CSAR, Lozano:aspdac25:BinCoDesign}.

FE does not aim to replace silicon-based platforms, but rather to complement them in domains where low cost, flexibility, and conformality are application-critical requirements.
Applications such as intelligent packaging, disposable medical devices, and distributed environmental sensing require electronic solutions that are not only functional but also flexible, lightweight, and economically viable~\cite{ Luo:smartpack, lee:flexiblepatch}.
\subsection{Binary Search ADC Architecture}\label{sec:overviewADC}

\begin{figure}
    \centering
    \includegraphics[width=1\linewidth]{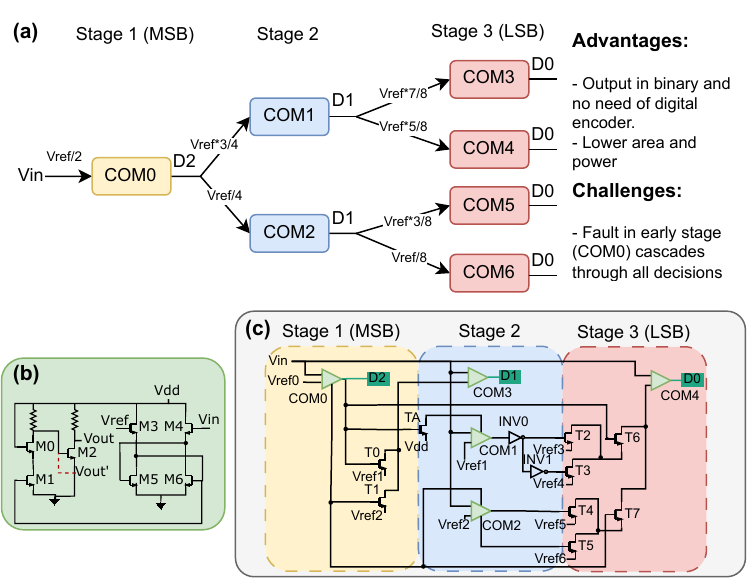}
    \caption{\purple{ (a) 3-bit Binary Search ADC conceptual architecture: cascaded comparators perform sequential decisions across three stages, producing binary output (D2, D1, D0) directly without digital encoding.} (b) Comparator in IGZO-TFTs used for the Binary Search ADC. (c) Binary Search ADC architecture.}
    \label{fig:Binary} \vspace{-2ex}
\end{figure}

Multiple ADC topologies have been realized in FE, each presenting distinct trade-offs. 
Flash converters~\cite{JAMSHIDIROUDBARI:2010:FlashFlexible} offer high speed through parallel comparison but demand large comparator arrays. 
Successive Approximation Register (SAR) architectures~\cite{Papadopoulos:C-2CSAR, Alkhalil:BioCAS:2022:FlexibleSAR} achieve good energy efficiency but require precision capacitor arrays challenging to implement in IGZO. 
Sigma-Delta converters~\cite{Garripoli:2017:DeltaSigmaADC, Wadhwa:ISCAS:2021:SigmaDeltaPrintedADC} provide excellent resolution but at the cost of complexity and oversampling requirements.

This work examines Binary Search ADC topology~\cite{Lozano:aspdac25:BinCoDesign}, which provides an attractive balance for FE applications operating at low to moderate sampling rates (sub-100 Hz) typical of biomedical and environmental sensing.

\textbf{Operational Principle:}
Binary Search conversion performs iterative voltage range subdivision. 
At each conversion step, a comparator determines whether the input voltage exceeds the midpoint of the current search window. 
Based on this decision, the search continues in either the upper or lower half of the remaining range. 
\purple{Figure~\ref{fig:Binary}(a) illustrates the complete decision tree for a 3-bit converter, showing how the input propagates through three cascade stages to produce the final digital code.}

\textbf{Architectural Advantages:}
The sequential nature enables significant resource savings compared to parallel topologies. 
A 3-bit Binary Search ADC requires only 3 comparison stages utilizing a total of 5 comparators (distributed across the decision tree), whereas an equivalent Flash converter would demand 7 parallel comparators plus a digital priority encoder. 
Beyond the reduced comparator count, the Binary Search architecture eliminates the need for complex digital encoding logic—the sequential decisions directly produce the binary output code. 
This dual benefit of fewer analog blocks and absent digital circuitry translates directly to decreased area and substantially lower static power consumption. 

\textbf{Comparator Design:}
Figure~\ref{fig:Binary}(b) illustrates the IGZO-based differential comparator employed in the baseline architecture. 
The circuit simultaneously generates both the comparison result and its logical complement, eliminating explicit inverter stages in the control path and reducing circuit complexity. 
The differential topology provides some inherent common-mode rejection, improving robustness to supply noise and substrate coupling.

\textbf{Implementation in N-Type Only Technology:}
\orange{The Binary Search ADC is particularly well-suited for unipolar transistor technologies including IGZO, ITO, GaN, and NCFETs, which lack complementary p-type devices. These technologies share common design constraints: }
The sequential operation tolerates the longer propagation delays inherent to large-geometry n-type transistors, as conversion speed requirements remain modest for biomedical applications. 
The reduced component count mitigates the impact of large device sizes on total chip area. 
\orange{While this work validates the methodology using IGZO-TFTs, the fault-aware design approach applies directly to other n-type only technologies facing similar defect density and process variation challenges.}
Figure~\ref{fig:Binary}(c) presents the \purple{transistor-level implementation of the} complete 3-bit Binary Search ADC architecture.

\subsection{Fault Simulation for Analog Circuits}\label{sec:related_work}

Simulating faults in analog circuits demands significant computational resources due to the extensive fault space and the absence of universally accepted behavioral-level fault models~\cite{BinuSurvey}. Although considerable research exists on analog fault simulation, we concentrate here on approaches that facilitate high-level fault modeling through behavioral abstractions extracted from transistor-level simulations.

Multiple mathematical representations have been investigated to approximate circuit responses under faulty conditions. Chebyshev and Newton interpolating polynomials have been employed to model nonlinear circuit characteristics within unified state-space frameworks~\cite{XiaChebyshev}. Unfortunately, these representations frequently fail to account for timing or phase perturbations induced by defects. Other approaches have attempted to correlate fault impacts with deviations in performance metrics~\cite{SubrahmaniyanAdaptive}, yet prove inadequate when faults fundamentally transform system functionality.

Approaches utilizing complex field representations~\cite{YangComplex} and system identification techniques~\cite{LuoSystemID} have been deployed for linear time-invariant topologies, characterizing modifications in transfer functions and impulse responses. These methodologies, however, remain confined to particular circuit categories and linear operating regimes. Symbolic extraction methods, exemplified in~\cite{FraccaroliAbstraction}, derive differential equations from signal flow representations but presume simplified device models that neglect nonlinear or non-ideal fault manifestations.

Frequency-domain techniques, including Hilbert transform analysis~\cite{LiuHilbert}, have been utilized to quantify both amplitude and phase variations. Statistical defect simulation frameworks employing weighted sampling~\cite{BhattacharyaDefect} and nonlinear autoregressive modeling with automated VHDL-AMS generation~\cite{FarooqNLARX} have also been demonstrated. These methods, while offering certain benefits, encounter obstacles such as prohibitive search spaces or insufficient exploitation of circuit topology.


The template-driven hierarchical methodology described in~\cite{aksoy2021hierarchical} constitutes a notable contribution in this domain. By refining behavioral representations through functional augmentation informed by transistor-level fault characterization, it achieves accurate fault simulation with substantially reduced computational burden. Validation was performed on mixed-signal architectures including Flash ADCs and Phase-Locked Loops.

Despite these advances, hierarchical fault simulation applied to FE technologies remains largely unexplored. The work in~\cite{PalAZSHZBT24} analyzed fault sensitivity in printed Multilayer Perceptron Classifiers, demonstrating that the distinctive attributes of FE, process-induced variability, unipolar device operation, and leakage-dominated static power, demand specialized modeling techniques.

In this work, we adapt and extend the hierarchical framework from~\cite{aksoy2021hierarchical} to examine fault modeling in flexible Binary Search ADCs. The sequential conversion architecture presents distinct fault propagation characteristics compared to parallel topologies, requiring tailored abstraction strategies to capture fault effects efficiently across the conversion hierarchy.

%% file: section/3_Methodology.tex
\section{Fault Injection Methodology} \label{sec:metho}

\begin{figure*}[t]
    \centering
    \includegraphics[width=1\linewidth]{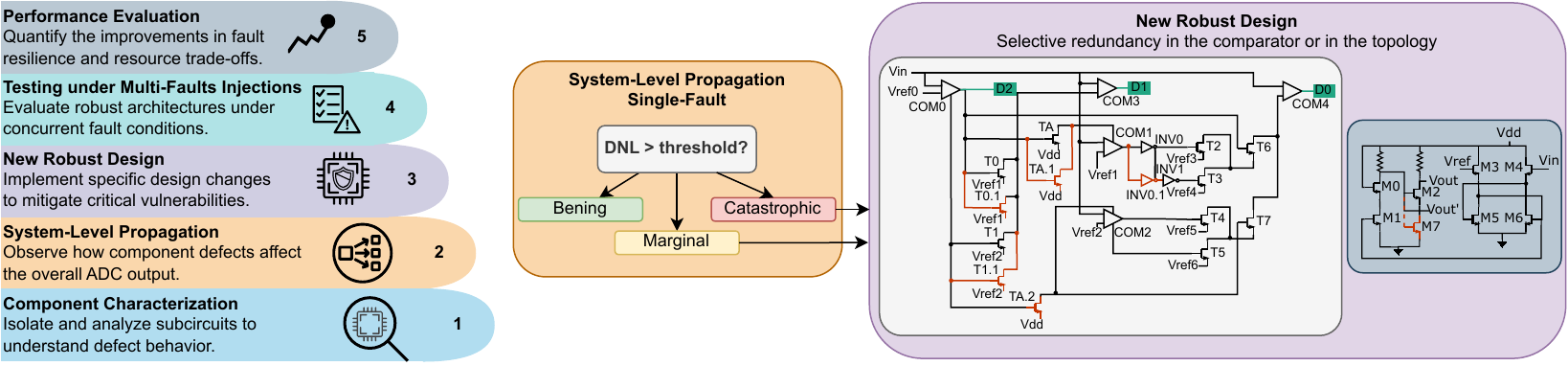}
    \caption{Fault injection workflow: subcircuit fault characterization builds behavioral models used in system-level simulation to identify vulnerabilities, guiding targeted design hardening validated through multi-fault stress testing.}
    \label{fig:hierarchical_flow} \vspace{-3ex}
\end{figure*}

\purple{The Binary Search ADC's cascaded decision structure introduces unique fault propagation characteristics distinct from parallel converter architectures, such as Flash ADC.
In Flash converters, individual comparator failures affect only specific output codes and the converter remains partially functional with degraded accuracy. 
In Binary Search architectures, however, a fault in an early conversion stage can misdirect the entire subsequent search process, potentially corrupting multiple output codes.
This hierarchical fault propagation mechanism—where early-stage faults cascade through later decisions—necessitates a methodology capable of capturing both localized defect behavior and system-level propagation patterns to identify critical vulnerability points.
Figure~\ref{fig:hierarchical_flow} illustrates the complete fault analysis workflow, progressing from transistor-level characterization through system-level assessment and culminating in architectural validation.
}

\subsection{Defect Modeling at Transistor Level}
Structural defects in IGZO-TFT circuits are modeled by injecting resistive parasitics into Spectre transistor-level netlists. 
Two mechanisms are considered: \textit{opens}, represented by high-impedance paths (\(250~\text{M}\Omega\)) at transistor terminals, and \textit{shorts}, represented by low-impedance bridges (\(10~\Omega\)) between circuit nodes. This modeling approach establishes the foundation for systematic fault injection in subsequent analyses.

\purple{
\subsection{Fault Library Generation at Subcircuit Level}
Using the transistor-level defect models, individual comparator subcircuits are extracted from the ADC hierarchy and subjected to exhaustive fault injection. Each transistor terminal—gate, drain, and source—is tested independently for both open and short conditions. 
Transient simulations capture the resulting behavioral deviations, constructing a fault library that maps specific defect locations to their functional impact on the conversion operation.
}

\subsection{System-Level Impact Assessment}

Faulty models substitute nominal instances in the complete Binary Search ADC netlist. 
Full analog-to-digital conversion sequences are simulated with representative input signals. The digital output code sequence is analyzed to compute Differential Nonlinearity (DNL) across the full-scale range.
Fault impact is classified into three categories based on DNL degradation severity:
\begin{itemize}
    \item \textbf{Catastrophic:} Complete converter malfunction or output codes stuck at incorrect 
    \item \textbf{Marginal:} Significant DNL degradation ($0.5 < DNL < 1$ LSB)  but conversion remains operational
    \item \textbf{Benign:} Minimal DNL impact ($<0.5$ LSB), functionally acceptable performance
\end{itemize}

This classification produces a statistical vulnerability profile revealing which transistors and circuit locations dominate fault sensitivity.






\subsection{Multiple Fault Scenarios}

Single-fault analysis identifies individual vulnerabilities but fails to capture defect clustering effects that occur in realistic manufacturing scenarios. Multi-fault test cases are constructed by strategically combining faults based on their individual classification:
\begin{itemize}
    \item Pairs of catastrophic faults (worst-case combinations)
    \item Catastrophic + marginal combinations (realistic clustered defects)
    \item Multiple marginal faults (accumulated degradation effects)
\end{itemize}

Both baseline and fault-resilient designs undergo these multi-fault stress tests. 
The validation determines whether the architectural improvements maintain acceptable DNL under concurrent defect conditions or whether fault interactions produce unexpected failure modes beyond those predicted by single-fault analysis.

\subsection{Performance Metric Extraction}

\purple{For each fault scenario, both single and multi-fault, quantitative metrics are extracted:}

\begin{itemize}
    \item \textbf{Fault coverage:} Percentage of injected faults producing acceptable DNL ($DNL<1$ LSB)
    \item \textbf{DNL statistics:} Distribution of linearity errors across the fault space, characterizing degradation severity
    \item \textbf{Hardware overhead:} Area increase and power consumption relative to baseline under both nominal and faulty operating conditions
\end{itemize}

This comprehensive dataset quantifies the resilience improvement achieved by each proposed design and characterizes the associated trade-offs in circuit resources.

\subsection{Automated Simulation Environment}

\begin{figure}[t]
    \centering
    \includegraphics[width=1\linewidth]{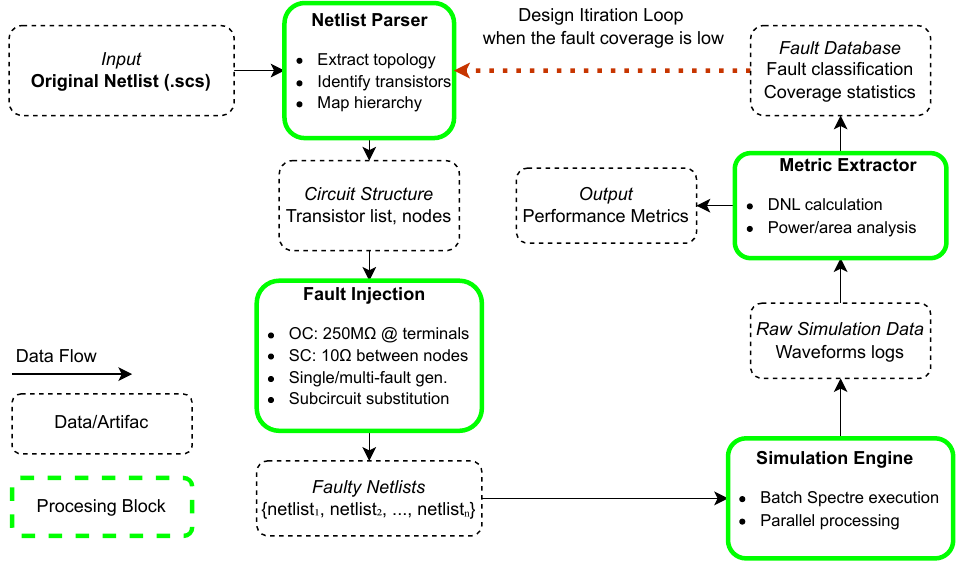}
    \caption{Automated fault simulation infrastructure: Python framework coordinates netlist manipulation, parametric fault insertion, batch execution, and metric extraction to enable comprehensive robustness evaluation.}
    \label{fig:automation_flow} \vspace{-3ex}
\end{figure}

Figure~\ref{fig:automation_flow} illustrates the proposed automated simulation environment developed to streamline fault analysis and design optimization. 
The entire workflow is managed by a custom Python framework that coordinates netlist manipulation, parametric fault injection, batch simulation, and metric extraction. 

The process begins with the parsing of Spectre-format netlists to extract the circuit topology, hierarchical subcircuit structure, and device instances. 
A dedicated fault injection engine then introduces resistive fault models at predefined transistor terminals, supporting both single-fault and multi-fault scenarios.
Once the faulty netlists are generated, the framework manages their parallel execution through Spectre and Ocean, substantially reducing simulation time. 
The resulting data are processed through an automated post-analysis pipeline that computes key metrics such as DNL, power, and area, followed by statistical aggregation across all fault distributions.

This integrated automation environment allows the systematic evaluation of thousands of fault configurations while maintaining full traceability of results, thus facilitating rapid and reproducible design iterations.

\section{Vulnerability-Aware Fault-Tolerant Design} \label{sec:fault_tolerant}

\purple{The fault sensitivity profile obtained from the hierarchical injection framework guides selective architectural improvements to enhance robustness.
Rather than applying uniform redundancy across the entire design, mitigation strategies target specific high-impact locations identified through the vulnerability analysis. Two complementary approaches are explored:}

\begin{itemize}
    \item \textbf{Topology Redundancy Strategy:} Components identified as catastrophic fault sources receive duplication with output voting or comparison mechanisms. This targeted approach adds minimal overhead compared to full redundancy while directly addressing the dominant failure modes revealed by sensitivity analysis.

    \item \textbf{Comparator Architecture Modification:} Comparators are redesigned to reduce dependency on fault-sensitive transistors. 
\end{itemize}

Alternative topologies distribute critical functionality across multiple signal paths, preventing single-point failures from compromising overall operation. 
This architectural diversity improves resilience without replicating the entire circuit.

\subsection{Fault-Tolerant Architecture Validation}

\purple{The proposed fault-tolerant designs undergo the complete fault injection methodology described in Section~\ref{sec:metho}, encompassing both single and multi-fault scenarios. This validation serves two purposes: first, to confirm that architectural improvements successfully mitigate the identified vulnerabilities; second, to assess whether fault interactions in the modified designs produce unexpected failure modes not evident in baseline analysis.}
\purple{Both baseline and fault-tolerant architectures are subjected to identical fault injection tests. Comparative analysis extracts:}
\begin{itemize}
    \item \textbf{Fault coverage improvement:} Change in percentage of acceptable fault scenarios relative to baseline
    \item \textbf{Hardware overhead:} Area increase and power consumption relative to baseline under both nominal and faulty operating conditions
    \item \textbf{Component-level resilience:} Per-subcircuit fault coverage to identify remaining vulnerabilities
\end{itemize}

%% file: section/4_Results.tex
\section{Experimental Setup and Results}\label{sec:results}

\subsection{Simulation Environment and Design Specifications}

\begin{table}
\centering
\caption{Component sizing specifications for the baseline Binary Search ADC comparator and resistive reference ladder.}
\label{tab:sizing}
\scalebox{1}{\input{tables/sizing}}
\end{table}

\textit{Technology Platform and Design Tools:}
All circuit simulations utilize Cadence Spectre with Pragmatic's second-generation FlexIC process design kit (PDK). A Python~3.9.18 automation layer orchestrates the complete simulation pipeline, managing netlist manipulation, fault injection, batch execution, and performance metric extraction.

\textit{ADC Resolution and Architectural Scope:}
This work validates the methodology on a 3-bit Binary Search ADC architecture. The hierarchical fault injection framework and fault-resilient strategies generalize to higher-resolution implementations, though the fault space complexity scales with bit count.

\textit{Operating Conditions:}
The converter operates at $V_{dd}$ 1V, with a 5Hz sinusoidal input signal spanning the full 1V range. 
Table~\ref{tab:sizing} details transistor dimensions and resistor values for the baseline design, maintaining consistency with the validated architecture in~\cite{Lozano:aspdac25:BinCoDesign}.
\orange{We employ dedicated thin-film resistors (200 $k\Omega/sq$ layer) provided by the FlexIC PDK. While resistor mismatch is inherently included in the PDK Monte Carlo models, the fault injection framework focuses on catastrophic open/short defects rather than parametric variation, as the differential comparator topology provides inherent tolerance to common-mode reference shifts.}

\textit{Fault Space Coverage:}
Each architectural variant (baseline and proposed designs) undergoes evaluation across more than 220 distinct fault configurations. 
This encompasses:
\begin{itemize}
    \item All single-fault scenarios: OC and SC at every transistor terminal
    \item Representative multi-fault combinations: catastrophic pairs, catastrophic+marginal, multiple marginal
\end{itemize}

Comprehensive coverage ensures that results reflect genuine architectural robustness rather than fortuitous resilience to a limited test subset.

\textit{Performance Metrics:}
Static linearity is quantified through DNL analysis computed from transition voltages in the digital output codes. 
DNL serves as the primary fault classification and design comparison metric, as it directly reflects conversion accuracy degradation under defect conditions.

\subsection{Baseline Fault Sensitivity Analysis}\label{sec:baseline}

\begin{figure*}
\centering
\includegraphics[width=0.85 \linewidth]{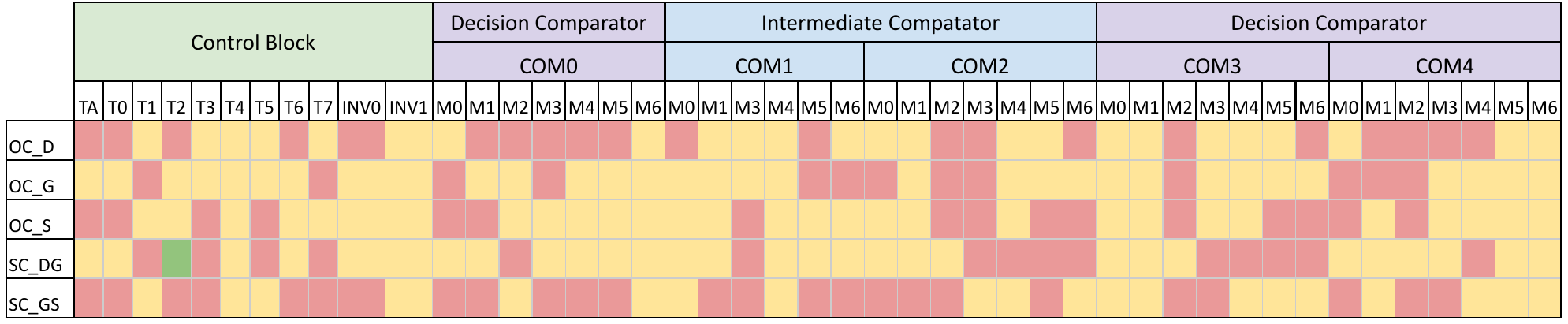}\vspace{-2ex}
\caption{Binary Search ADC fault sensitivity heat map under single-fault injection. Red: catastrophic; yellow: marginal; green: benign.}
\label{tab:results_bin} \vspace{-3ex}
\end{figure*}

\begin{table}
\caption{Comparison of comparator configurations in terms of fault coverage, area, and power.}
\label{tab:results_faults}
\scalebox{1}{\input{tables/resultsFaults}}
\end{table}

Figure~\ref{tab:results_bin} visualizes RMS DNL deviations for transistor-level single-fault injections in the baseline 3-bit Binary Search ADC. 
\textit{Fault effects are highly non-uniform:} catastrophic failures cluster in a small subset of transistors, notably in comparator input differential pairs and early-stage output stages, while the majority of faults produce marginal deviations.

\purple{The baseline analysis reveals several important characteristics. 
First, a small number of transistors dominate catastrophic failure modes, suggesting that \textit{selective mitigation is more effective and area-efficient than uniform redundancy.} 
Second, \textit{the hierarchical architecture amplifies early-stage faults}—defects in the MSB conversion stage cascade through subsequent stages, causing widespread output corruption. 
This hierarchical propagation highlights the importance of prioritizing robustness in early-stage components. 
Finally, the limited baseline coverage (60\% single-fault, 34\% multi-fault) indicates that concurrent defects significantly exacerbate reliability risk, a realistic concern in FE given elevated defect densities.}

These observations demonstrate that both fault location and hierarchical propagation patterns are essential considerations for targeted, efficient fault-tolerant design.

\subsection{Fault-Tolerant Architecture Evaluation}\label{sec:hardened}

Based on the baseline vulnerability analysis, two fault-resilient architectures were developed:

\purple{\textbf{Selective Front-End Redundancy (SFR):}} Implements selective redundancy at high-sensitivity transistors identified after the fault injection. Including input amplification stages (TA.1, TA.2), pass-gate stages (T0.1, T1.1), and the first inverter stage (INV0). Redundant devices maintain the original sizing (W = 2~\textmu m, L = 600~nm).

\purple{\textbf{Extended Comparator-Level Redundancy (ECLR):} } Extends SRF with additional output-stage redundancy and an extra comparator transistor (M7) in COM0, COM3, and COM4, targeting the comparators contributing most to catastrophic outcomes.

\begin{figure}
\centering
\includegraphics[width=1\linewidth]{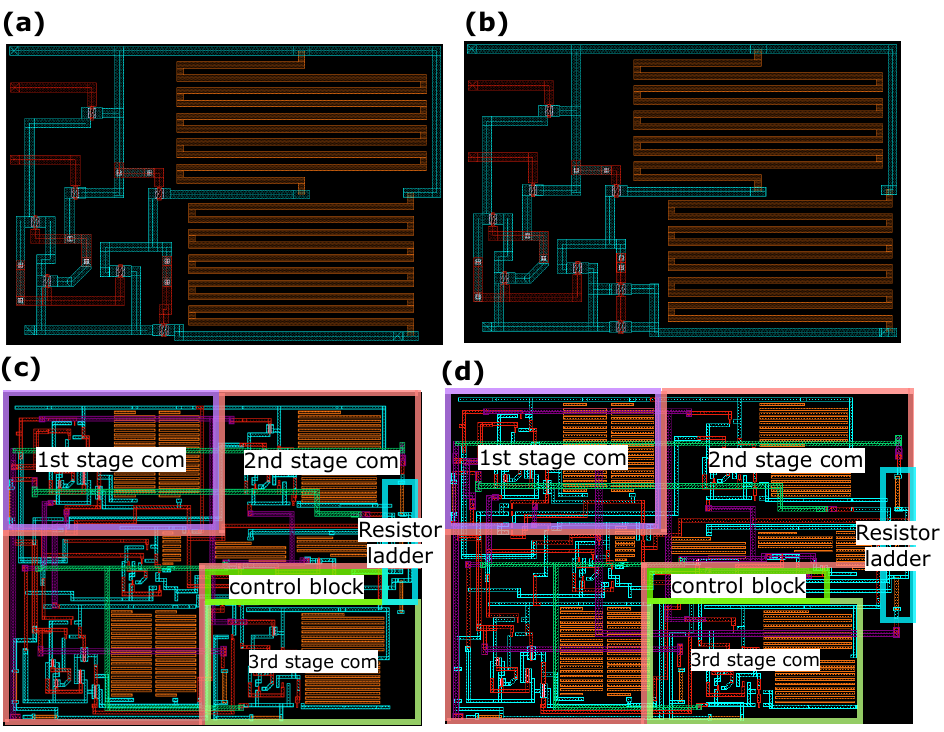}
\caption{Physical layout comparison: (a) baseline COM3 comparator (b) fault-resilient COM3 with extended redundancy (c) complete baseline 3-bit Binary Search ADC 
(d) fault-resilient design ECLR.} 
\label{fig:layouts} \vspace{-3ex}
\end{figure}

Figure~\ref{fig:hierarchical_flow} illustrates both architectures with redundancy locations highlighted, while Figure~\ref{fig:layouts} presents the physical layouts.

\subsubsection{System-Level Coverage Results}

Table~\ref{tab:results_faults} summarizes fault coverage for all designs under single and multi-fault injection. SFR achieves 88.9\% single-fault and 73.2\% multi-fault coverage, while ECLR reaches 92\% and 77.3\%, respectively, compared to the 60\%/34\% baseline.

\begin{table}[t]
\centering
\caption{Multi-fault coverage by circuit component.}
\label{tab:component_coverage}
\begin{scriptsize}
\begin{threeparttable}
\begin{tblr}{
  colspec = {Q[12] Q[8] Q[8] Q[8]},
  hline{1,2,8} = {-}{1.2pt}, 
  hline{3-7} = {-}{0.4pt},
  row{1} = {c, font=\bfseries},
  row{2-7} = {c}
}
\textbf{Component} & \textbf{Baseline} & \textbf{SFR} & \textbf{ECLR} \\
Control Block & 51.52\% & 51.52\% & 51.52\% \\
COM0 & 54.55\% & 63.64\% & 72.73\% \\
COM1 & 17.65\% & 82.35\% & 94.12\% \\
COM2 & 35.00\% & 90.00\% & 90.00\% \\
COM3 & 0.00\% & 100.00\% & 100.00\% \\
COM4 & 0.00\% & 83.33\% & 100.00\% \\
\end{tblr}
\end{threeparttable}
\end{scriptsize}\vspace{-3ex}
\end{table}

\subsubsection{Component-Level Analysis}

Table~\ref{tab:component_coverage} presents per-component multi-fault coverage. Output comparators COM3 and COM4, completely vulnerable in baseline (0\%), reach 100\% coverage in the proposed designs. Intermediate stages show varying improvement patterns: COM1 increases from 17.65\% to 94.12\%, while COM2 reaches 90\% in both SFR and ECLR. The Control Block remains at 51.52\% across all variants, and further system-level improvements would require addressing vulnerabilities in this control logic block.

The SFR-to-ECLR progression yields gains primarily in COM0 (+9.09\%), COM1 (+11.77\%), and COM4 (+16.67\%), while COM2 and COM3 show saturation, indicating diminishing returns from additional redundancy in these stages.

\subsubsection{Multi-Fault Interaction Analysis}

\blue{The substantial improvement in multi-fault coverage—from 34\% baseline to 77.3\% in ECLR—reflects the proposed architectures' ability to mitigate fault interaction effects. 
Analysis of specific multiple fault combinations reveals how the selective redundancy strategy prevents individual defect classifications from compounding into system failure, and representative examples illustrate these interaction patterns: When combining COM1\_M0\_OC\_G (marginal in baseline single-fault analysis) with COM4\_M4\_OC\_D (catastrophic in baseline), the unmodified design produces catastrophic DNL degradation under multi-fault simulation. Both SFR and ECLR reduce this to marginal classification, demonstrating that the \textit{architectural improvements successfully contain fault propagation}, even when a single defect causes catastrophic failure.}

\blue{Similarly, pairing two individually catastrophic faults—COM3\_M3\_OC\_S and COM4\_M4\_OC\_D—produces catastrophic system failure in the baseline. The fault-resilient designs again limit degradation to marginal DNL, confirming that \textit{selective redundancy at output stages prevents worst-case fault combinations from causing complete converter malfunction.}}

\blue{The architectural improvements can also reduce marginal fault accumulation. Combining COM3\_M1\_OC\_D (marginal) with COM2\_M2\_SC\_GS (marginal) yields marginal DNL in the baseline but benign classification in both SFR and ECLR. This downgrading demonstrates that the \textit{proposed designs not only prevent catastrophic failures but also improve tolerance to accumulated moderate degradations.}}


\subsection{Design Trade-off Analysis}\label{sec:tradeoff}

Physical layout implementation (Figure~\ref{fig:layouts}) confirms the area overhead: ECLR occupies 58,615~\textmu m$^2$ compared to the 56,259~\textmu m$^2$ baseline, representing a 4.2\% increase. 
\orange{This overhead accounts for the complete physical implementation, including routing complexity associated with redundant transistors; post-layout simulations validate that the fault coverage improvements are maintained with extracted parasitics.}
Combined with 5.9\% power overhead, ECLR achieves 92\% single-fault coverage and 77.3\% multi-fault coverage—a 32 and 43.3\% improvement, respectively.

SFR achieves 88.9\% single-fault and 73.2\% multi-fault coverage with slightly lower overhead (estimated 4.2\% area based on layout, 6.0\% power), indicating diminishing returns: the additional 3.1\% single-fault coverage observed from SFR to ECLR is obtained without any increase in area, as the comparator redesign primarily involves redistributing existing layout resources. 
The added inverter occupies minimal space, and the size of individual transistors remains negligible compared to that of resistive components.
For comparison, traditional triple modular redundancy (TMR) typically requires ~200\% area overhead, and dual redundancy ~100\%, to achieve comparable fault coverage. The selective approach achieves similar resilience at 2-3\% of the resource cost by targeting only vulnerability-critical locations identified through hierarchical analysis.

%% file: tables/sizing.tex
\begin{threeparttable}
\resizebox{\linewidth}{!}{%
\centering
\setlength{\arrayrulewidth}{0.4pt} 
\begin{tblr}{
  colspec = {Q[30] Q[25] Q[30]}, 
  vline{2,3} = {-}{0.4pt},
  hline{1,2,14} = {1.2pt}, 
  hline{6,8,11,13} = {-}{0.4pt}, 
  hline{3,4,5,7,9,10,11,12} = {2,3}{0.4pt}, 
}
\textbf{Component} & \textbf{Element} & \SetCell[c=1]{c} \textbf{Size} \\ 

\textbf{COM0\&COM2} & M0, M1, M2 & W= 8$\mu$m L= 600nm \\
 & M3, M4 & W= 2$\mu$m L= 600nm \\
 & M5, M6 & W= 5$\mu$m L= 600nm \\
 & R & r = 71$M\Omega$\\

\textbf{COM3\&COM4} & M0:M6 & W= 2$\mu$m L= 600nm \\
 & R & r = 71$M\Omega$\\
 
\textbf{COM1}$^1$ & M0:M4 & W= 2$\mu$m L= 600nm \\
 & M5, M6 & W= 5$\mu$m L= 600nm \\
 & R & r = 11.37$M\Omega$\\

\textbf{INV0\&INV1} & M & W= 3$\mu$m L= 600nm \\
 & R & r = 26$M\Omega$\\
\SetCell[c=2]{c} \textbf{T0:T7\&TA0 (Refereed as Control Block)} & & W= 2$\mu$m L= 600nm \\ 
\end{tblr}

}
\begin{tablenotes}\footnotesize
    \item[]$^1$COM1 only has Vout’ as output, second R and M2 are removed. \purple{The schematics of the comparator and the ADC are in Figure~\ref{fig:Binary} (b).} \vspace{-3ex}
\end{tablenotes}
\end{threeparttable}

%% file: tables/resultsFaults.tex
\begin{scriptsize}
\begin{threeparttable}
\centering
\resizebox{\linewidth}{!}{%
\begin{tblr}{
  colspec = {Q[4.5] Q[11.5] Q[11] Q[10.5] Q[8]},
  hline{1,2,5} = {-}{1.2pt}, 
  hline{3,4} = {-}{0.4pt},
  row{1} = {c, font=\bfseries},
  row{2-4} = {c}
}
\textbf{Design} 
& \textbf{Single Injection Coverage*} 
& \textbf{Multi Injection Coverage*} 
& \textbf{Area (\textmu m$^2$)} 
& \textbf{Power (nW)} \\

Baseline & 60\% & 34\% &56260&334.1\\
SRF &88.9\%&73.2\%&58615 (+4.2\%)&354 (+6\%)\\
ECLR &\textbf{92\%}&\textbf{77.3\%}&58615 (+4.2\%)&353.8 (+5.9\%)\\
\end{tblr}
}
\begin{tablenotes}
\footnotesize
\item *The coverage comes as the percentage of the codes with DNL$<$1. \vspace{-4ex}
\end{tablenotes}
\end{threeparttable} 
\end{scriptsize}

%% file: section/5_Conclusion.tex
\section{Conclusion}\label{sec:conclusion}

Flexible electronics face elevated defect densities and device variability, posing significant reliability challenges for analog circuits such as Analog-to-digital converters (ADCs). 
This work presents a fault-aware design methodology that combines transistor-level defect characterization with system-level propagation analysis, enabling efficient evaluation of both single and concurrent faults while capturing IGZO-specific failure mechanisms.


\purple{Applied to Binary Search ADCs—particularly attractive for resource-constrained wearable applications due to their reduced component count and elimination of digital encoding logic—the framework reveals that fault vulnerability concentrates in early conversion stages. 
Leveraging this insight, we develop fault-tolerant designs employing selective redundancy and topology modifications targeted at these critical early-stage comparisons.}
Experimental validation demonstrates substantial robustness improvements: single-fault coverage rises from 60\% to 92\%, and multi-fault coverage from 34\% to 77.3\%, with only 4.2\% area and 6\% power overhead. 
These results demonstrate that systematic fault sensitivity analysis enables targeted architectural improvements that effectively enhance reliability in flexible analog circuits without prohibitive resource costs. 